\documentclass[footinbib,twocolumn,showpacs,amsmath,amstex,amssymb,mathfonts,superscriptaddress,prl]{revtex4}
\usepackage{graphicx}
\usepackage{color}
\usepackage{bm}

\usepackage{graphicx}
\usepackage{color}
\usepackage{bm}
\usepackage{amsmath}
\usepackage{amssymb}
\usepackage{amsthm}
\usepackage{amsfonts}
\usepackage{multirow}
\usepackage{makecell}
\usepackage{float}
\usepackage{comment}
\usepackage{enumitem}
\usepackage{verbatim}
\usepackage{mathtools}

\usepackage{bbm}

\begin{document}

\title{Equivalence of Chern bands and Landau levels from projected Interactions}
\author{Bo Yang} 
\affiliation{Division of Physics and Applied Physics, Nanyang Technological University, Singapore 637371.}

\date{\today}
\begin{abstract}
We introduce a universal formulation of the generalized real space interactions with translational invariance, when projected into a single Landau level, can be equivalent to density-density interaction projected into any Chern bands (e.g. Landau levels, continuous moire Chern bands and discrete lattice Chern bands). By constructing a complete basis of generalized pseudopotentials, we define the projected interaction range that can be used to characterize important features of different Chern bands relevant to the robustness of interacting topological phases. The analytical construction allows us to study general Chern bands beyond the torus geometry, for example on the disk or spherical geometry where the edge dynamics and curvature effects can be explored. For moire systems (and lattice Chern bands with large incompressibility gap) one can see transparently the topological Hall viscosity for the fractional states are identical to those in the Landau levels. The physical unprojected interaction also provides a natural embedding for lattice systems, shedding light on the concerns of relying on quantum geometric tensor (varying with lattice embedding in real space) to understand the properties of interacting phases of matters.
\end{abstract}

\maketitle 

Two-dimensional (2D) topological bands with non-zero Chern number play one of the central roles in understanding exotic phases of matters and the novel engineering of low-dimensional quantum materials \cite{hasan2010colloquium, bergholtz2013topological, NiuReview2016}. From the single particle perspective, these bands have non-zero and generally non-uniform Berry curvature in momentum space, leading to anomalous transport properties at the fermi surface including both the linear and higher order Hall effect \cite{Klitzing1980, ma2019observation,PhysRevLett.115.216806,PhysRevLett.121.266601, kang2019nonlinear,du2021nonlinear}. For partially filled bands in the presence of strong electron-electron interaction, highly entangled topological phases emerge. The typical example is the fractional quantum Hall (FQH) effect, hosting fascinating collective excitations from fractionally charged anyons to neutral gravitons \cite{Tsui1982, laughlin1983anomalous,Yang2012,PhysRevB.87.245132,liou2019chiral, wang2021analytic,Yuzhu2023, Liang2024,wang2025dynamics}. Chern bands in different quantum materials can have drastically different physical properties. While the single particle physics is relatively well understood, it becomes very challenging to understand the fundamental differences and similarities between different Chern bands in the presence of strong interactions and various competing strongly correlated phases\cite{Xie2021,PhysRevLett.128.156401,PhysRevLett.132.036501,lu2024fractionalquantumanomaloushall}.

The Landau levels (LL), realised from free two-dimensional electron gas subject to a strong perpendicular magnetic field, is the simplest family of 2D Chern bands with zero dispersion (constant kinetic energy) and uniform quantum geometric tensor or QGT (i.e. the Berry curvature and the Fubini-Study metric\cite{berry1984quantal,PhysRevLett.51.2167,Roy2014}). Much of the understandings of the more complicated Chern bands with non-uniform QGT and non-trivial boundary conditions are drawn from the analogies to the LLs. These include the ideal flat band conditions\cite{Wang2021}, vortexability\cite{Ledwith2023} and the generalized LL formalism \cite{liu2024theorygeneralizedlandaulevels}. The main strategy is to compare these Chern bands with the LLs at the wavefunction level, by analyzing the single particle geometric properties from the QGT, and exploring conditions for the holomorphicity of the single particle states as is the case in the lowest LL (LLL). Both notions can be generalized to the lowest $N$ LLs. Numerical evidence shows that at least for certain cases, better approximation of LLs leads to more robust fractional quantum anomalous Hall states analogous to the FQH states in the LLs, especially in Moire systems described by the effective continuous model \cite{Wang2021,LIU2024515}.

This is however not true in general. Even for the simplest ideal flat bands with Coulomb interaction, the Laughlin phases can only exist for some of them, when the fluctuation of the Berry curvature is not too large\cite{Wang2021}. It is important to note that the physics of fractional topological states is determined by the short range part of the real space interaction (i.e. large momentum transfer) \cite{haldane1983hierarchy}. In contrast, the QGT is only the leading terms in the small momentum transfer expansion of the form factors, thus missing a lot of information. It is thus likely not fundamental and does not capture the rich physics of especially the gapped excitations or gapless phases in generic topological bands. In addition Chern bands can be realized either in continuous systems such as the LLs and the Moire systems, or in discrete lattice systems such as the Haldane and the Checkerboard models \cite{haldane1988model, sun2011nearly, Sheng2011, neupert2011fractional}. A unified language to understand all such Chern bands is thus desirable. Conceptually for interacting systems, it is also useful to characterize the Chern bands entirely from the nature of interaction within a single band, without explicitly referring to the single particle band geometry such as the QGT\cite{PhysRevB.102.165148}. 

In this Letter, we give a systematic characterization of any Chern bands (be it LLs, Moire Chern bands or lattice Chern bands) entirely from the perspective of \emph{real space interactions}. Starting with the bare density-density interaction
\begin{eqnarray}\label{hbare}
\hat H_{\text{bare}}=\int d^2r_1d^2r_2V_{\vec r_1-\vec r_2}\hat c^\dagger_{\vec r_1}\hat c^\dagger_{\vec r_2}\hat c_{\vec r_1}\hat c_{\vec r_2}
\end{eqnarray}
we study its effective form in different sub-Hilbert spaces, and show that all Chern bands can be understood as Landau levels with a \emph{generalized real space interaction}, and such interaction can be expanded in the complete basis of generalized pseudopotentials with momentum dependent coefficients. Using the shortest range real space interaction\cite{kivelson1985exact} in Eq.(\ref{hbare}), useful properties of many-body quantum fluids in the Chern bands can be either analytically or numerically computed by the introduction of the concept of projected interaction range, an effective interaction length scale within a truncated sub-Hilbert space. Armed with these analytical tools, we show that even in periodic systems, the low energy subspaces (i.e. ground state and gapless edge/quasihole excitations) of topological phases can be ``identical" to those in the LLs with emergent full rotational symmetry. Beyond LLs, new physics in lattice and Moire Chern bands comes from explicit breaking of rotational symmetry, which is only relevant to gapped excitations, gapless quantum fluids or close to the phase transitions. Moreover, the formulation proposed here allows us to analyze properties of any Chern bands on disk or spherical geometries. This is useful for studying the edge dynamics and Hall viscosity of the topological phases realized in these bands.  

{\it The general interaction in real space--} Let us start with a simple formulation and set up the notations, by focusing on interactions with translational invariance in the two-dimensional continuum. The most general interaction in real space is given as follows:
\begin{eqnarray}\label{generalr}
\hat H=\int d^2r_1d^2r_2d^2r_3d^2r_4V_{\vec r_1-\vec r_2,\vec r_3,\vec r_4}\hat c^\dagger_{\vec r_1+\vec r_3}\hat c^\dagger_{\vec r_2+\vec r_4}\hat c_{\vec r_1}\hat c_{\vec r_2}
\end{eqnarray}
where $\hat c^\dagger_{\vec r}$ and $\hat c_{\vec r}$ are the creation and annihilation operators of (quasi)particles. If $\hat c^\dagger_{\vec r}$ creates an electron at position $\vec r$ in the full Hilbert space, then we must have $V_{\vec r_1-\vec r_2,\vec r_3,\vec r_4}\sim\delta_{\vec r_3,0}\delta_{\vec r_4,0}$, going back to Eq.(\ref{hbare}) with the density-density interaction. However for electrons confined within a sub-Hilbert space, e.g. a single band, Eq.(\ref{generalr}) is physically relevant. This is even true for any tight-binding model with $\hat c_{i\alpha}^\dagger$ creating an electron on the $i^{\text{th}}$ unit cell with the sub-lattice index $\alpha$ at position $\vec r_{i\alpha}$. Note the quantum state localized at $\vec r_{i\alpha}$ is a localized orbital with localization length $\ell_a$. A density-density interaction in the orbital basis (e.g. the Hubbard nearest neighbour interaction $V_{ij}^{\alpha\beta}\hat n_{i\alpha}\hat n_{j\beta}$ with $\hat n_{i\alpha}=\hat c^\dagger_{i\alpha}\hat c_{i\alpha}$) is only correct in the limit $\ell_a\to 0$. For finite $\ell_a$ any physical interaction is a discrete version of Eq.(\ref{generalr})\cite{footnote}.

In the momentum space Eq.(\ref{generalr}) is equivalent to 
\begin{eqnarray}\label{generalk}
\hat H=\int d^2k_1d^2k_2d^2q\tilde V_{\vec q,\vec k_1,\vec k_2}\hat d^\dagger_{\vec k_1+\vec q}\hat d^\dagger_{\vec k_2-\vec q}\hat d_{\vec k_1}\hat d_{\vec k_2}
\end{eqnarray}
where $\tilde V_{\vec k_1,\vec k_2,\vec k_3}$ is the fourier mode of $V_{\vec r_1,\vec r_2,\vec r_3}$ and $\hat d_{\vec k}\sim\int d^2re^{i\vec k\cdot \vec r}\hat c_{\vec r}$ on an infinite plane. If the two-dimensional manifold is on the torus with periodic boundary condition (PBC), then we have:
\begin{eqnarray}\label{generalkt}
&&\hat H=\sum_{\vec k_1\vec k_2\vec k_3\vec k_4}\tilde V_{\vec k_1\vec k_2\vec k_3\vec k_4}\hat d^\dagger_{\vec k_3}\hat d^\dagger_{\vec k_4}\hat d_{\vec k_1}\hat d_{\vec k_2}\\
&&\tilde V_{\vec k_1\vec k_2\vec k_3\vec k_4}=\sum_{\vec b}\tilde V_{\vec k_1-\vec k_4-\vec b,\vec k_1,\vec k_2}\label{vkt}
\end{eqnarray}
with the constraint $\vec k_1+\vec k_2=\vec k_3+\vec k_4+\delta \vec b$ due to discrete translational invariance, where $\vec k_i$ are within the first Brillouin zone (BZ) of the momentum space, and $\vec b,\delta \vec b$ are the reciprocal vectors. Note on this compact manifold in principle we need to set the total number of single particle states as $N_o$, which is also the total number of $\vec k$ points in the BZ. For a generic model $N_o$ is arbitrary, and for a specific physical system it is determined by the system length scale (e.g. the magnetic length or the lattice constant).

{\it General Chern bands as perturbed Landau levels--}Let us first start with LLs as simple but non-trivial examples. To compare with discussions later we always use the single particle Bloch states labeled by two linear momenta $|\vec k\rangle=|k_x,k_y\rangle$ defined by the magnetic translation operators. On an infinite plane the two-body interaction Hamiltonian projected into the LL is given by Eq.(\ref{generalk}) with the following:
\begin{eqnarray}\label{llk}
\tilde V_{\vec q,\vec k_1,\vec k_2}=\tilde V_{\vec q}\cdot \left(\mathcal L_N\left(q^2/2\right)\right)^2e^{-\frac{1}{2}q^2}\cdot e^{\frac{i}{2}\vec q\times\left(\vec k_1-\vec k_2\right)}
\end{eqnarray}
where $q=|\vec q|$, $\mathcal L_N\left(x\right)$ is the Laguerre polynomial and $N$ is the LL index; $\tilde V_{\vec q}$ is the fourier modes of the bare interaction in Eq.(\ref{hbare}), and the second part of Eq.(\ref{llk}) are the form factor from the single particle wavefunction in the $N^{\text{th}}$ LL. Here we also set the magnetic length $\ell_B=\sqrt{\hbar/eB}$ as unity, where $e$ is the electron charge and $B$ is the perpendicular magnetic field. The last part of Eq.(\ref{llk}) comes from the uniform magnetic field (i.e. uniform Berry curvature).

The single particle states of the lattice or Moire Chern bands are Bloch states just analogous to those in the LLs on the torus without quasi-periodicity, and with the reciprocal vectors $\vec b$ specifying the lattice type \cite{Wang2021}. Starting with a bare interaction $\tilde V_q$ in the full Hilbert space, its projection into a generic Chern bands with single particle states $|\vec k\rangle$ on the torus is given by Eq.(\ref{generalkt}) and Eq.(\ref{vkt}) with the following:
 \begin{eqnarray}\label{vktc}
 \tilde V_{\vec q\vec k_1\vec k_2}=\tilde V_q\langle u_{\vec k_4}|u_{\vec k_1}\rangle\langle u_{\vec k_3}|u_{\vec k_2}\rangle=\tilde V_q\mathcal F_{\vec k_1\vec k_2\vec k_3\vec k_4}
 \end{eqnarray}
where in real space $\langle \vec r|\vec k\rangle\sim e^{i\vec k\cdot\vec r}\langle \vec r|u_{\vec k}\rangle$ and we naturally require $\tilde V_{\vec q\vec k_1\vec k_2}=\tilde V_{\vec q,\vec k_1+\vec b,\vec k_2}=\tilde V_{\vec q,\vec k_1,\vec k_2+\vec b}$. The physical density-density interaction $\tilde V_q$ does not depend on the momenta of the states. In lattice models if the onsite or nearest neighbour interaction is used, $\tilde V_q$ is the fourier transforms of finite range step functions in real space. For the LLL, the form factor $\mathcal F_{\vec k_1\vec k_2\vec k_3\vec k_4}$ is given by\cite{Wang2021}
 \begin{eqnarray}\label{flll}
 \mathcal F^{\text{LLL}}_{\vec k_1\vec k_2\vec k_3\vec k_4}=e^{\frac{i}{2}\left(\left(\vec q-\vec b)\times\left(\vec k_1-\vec k_2\right)-2\vec q\times \vec b+\left(\vec k_2+\vec q\right)\times\delta \vec b\right)\right)}e^{-\frac{1}{2}q^2}
 \end{eqnarray}
 which is the torus version of Eq.(\ref{llk}). Any interactions in Moire or lattice Chern bands can be written in the form of Eq.(\ref{vktc}), thus their numerical computations can be done within a unified framework\cite{footnote}. 
 
Given the form of Eq.(\ref{flll}) we can rewrite Eq.(\ref{vktc}) as follows
\begin{eqnarray}\label{cb}
&&\tilde V_{\vec q\vec k_1\vec k_2}=\bar V_{\vec q\vec k_1\vec k_2}\tilde{\mathcal F}^{\text{LLL}}_{\vec k_1\vec k_2\vec k_3\vec k_4}\\
&&\bar V_{\vec q\vec k_1\vec k_2}=\tilde V_q\mathcal F_{\vec k_1\vec k_2\vec k_3\vec k_4}\left(\mathcal F^{\text{LLL}}_{\vec k_1\vec k_2\vec k_3\vec k_4}\right)^{-1}
\end{eqnarray}
Thus we can understand the interaction in any Chern band in two equivalent ways: the projection of Eq.(\ref{hbare}) into the Chern band, or the projection of an effective bare interaction of Eq.(\ref{generalr}) into the LLL, where $\bar V_{\vec q\vec k_1\vec k_2}$ is the fourier component of $V_{\vec r_1-\vec r_2,\vec r_3,\vec r_4}$. It is also important to note the length scale in Eq.(\ref{cb}) is set to be unity for convenience. This length scale is $\ell=\sqrt{\mathcal A/(2\pi N_o)}$, where $N_o$ is the number of states in the band and $\mathcal A$ is the surface area of the two-dimensional sample: it is the magnetic length in LLs and the lattice constant in any Chern insulator.

To put the LLs and the Chern bands on more firm equal footing, we can expand the projected effective interaction $\bar V_{\vec q\vec k_1\vec k_2}$ with an orthonormal set of generalized pseudopotential basis as follows:
\begin{eqnarray}\label{gppg}
&&\tilde V^{\text{LL}}_{\vec q\vec k_1\vec k_2}=\sum_{m,n}\lambda_{m,n,\vec k_1,\vec k_2}\tilde V_{m,n}\left(\bm q,\bm q^*\right)\\
&&\tilde V_{m,n}\left(\bm q,\bm q^*\right)\sim\left(\bm q\right)^n\mathcal L_m^n\left(q^2\right)e^{-\frac{1}{2}q^2}
\end{eqnarray}
where $\mathcal L_m^n\left(x\right)$ is the generalized Laguerre polynomials. Comparing with the well-known pseudopotential expansions in the LL\cite{haldane1983hierarchy,yang2017generalized,footnote}, which is a special case of Eq.(\ref{gppg}), there are two crucial aspects. Firstly there is the $\vec k_1,\vec k_2$ dependence on the expansion coefficients because $\vec r_3, \vec r_4$ in Eq.(\ref{generalr}) can be nonzero. Secondly the anisotropic pseudopotentials\cite{yang2017generalized} are involved even if $V(\vec r)$ in Eq.(\ref{hbare}) is rotationally invariant, due to the breaking of rotational symmetry by the lattice periodicity and thus the single particle form factors. In the complete orthonormal basis $\tilde V_{m,n}(\bm q,\bm q^*)$, the integer $n$ can be either positive or negative, and it maps a two-particle state from relative angular momentum $m$ to $m+n>0$. For $n=0$ we have the usual rotationally invariant Haldane pseudopotentials $\tilde V_m(q^2)=\tilde V_{m,0}(\bm q,\bm q^*)$. The generalized pseudopotential basis thus offers a universal language and analytic tool for understanding the interactions in all Chern bands.
 
 {\it Projected interaction range for ideal Chern bands--}
To exploit this analytic tool, we first show while different LLs are topologically equivalent (with the same Chern number), there are also qualitative differences not discussed in the literature before, that can be revealed from the nature of interactions. This can be illustrated by choosing the shortest possible real space interaction $V(\vec r_1-\vec r_2)=\delta^2\left(\vec r_1-\vec r_2\right)$ as the \emph{model interaction} for bosons to characterize the nature of the Chern bands with uniform QGT and rotational invariance (i.e. linear combination of LLs)\cite{footnote1}.  With the corresponding $\tilde V_q=1$ in Eq.(\ref{llk}), we have the following pseudopotential expansion with \emph{finite} number of terms:
 \begin{eqnarray}\label{formexpand}
 \tilde V_q\left(\mathcal L_N\left(q^2/2\right)\right)^2e^{-\frac{1}{2}q^2}=\sum_{m=0}^{N} \lambda_m\tilde V_{2m}(q^2)
 \end{eqnarray}
with $\lambda_m>0$ being readily computable, and crucially $\lambda_m=0$ for $m>N$, which can be proven by simply looking at the power of $q$. Thus with the shortest range interaction in real space, different LLs can be characterized by \emph{different number of pseudopotentials} present in the effective interaction. The higher the LLs, more pseudopotentials will be present and the projected interaction is effectively more ``long ranged". In particular, this implies the exact bosonic Laughlin state at $\nu=1/2$ can only be realised in the LLL, the exact bosonic Laughlin state at $\nu=1/4$ can only be realised in the LLL and the first LL, etc. 

We can thus define $\bm r_c$ as the projected interaction range of the Chern bands. Let $\mathcal F(q^2)$ be the generic form factor and since $\tilde V_q=1$, the pseudopotential expansion of the effective interaction is given by
\begin{eqnarray}
\mathcal F(q^2)e^{-\frac{1}{2}q^2}=\sum_{m=0}^{\bm r_c}\lambda_m\tilde V_{2m}(q^2)
\end{eqnarray} 
and $\bm r_c$ is the integer given by the largest $m$ of $\lambda_{m}$ that is non-zero such that $\lambda_{m>\bm r_c}=0$, indicating that the Chern band is a linear combination of the first $\bm r_c$ LLs. From $\bm r_c$ one can then have a very good indication of how easily the Laughlin states of different filling factors can be realized even with realistic interactions in such Chern bands. For the Chern band to be \emph{identical} to one of the LLs, we can treat $\lambda_m\sim\vec \lambda$ as a vector, and every LL is identified by a unique vector that is explicitly computed from Eq.(\ref{formexpand}).

We now proceed to characterize general Chern bands in analogy to LLs with generalized pseudopotentials. The model interaction with $\tilde V_q=1$ gives Eq.(\ref{cb}) corresponding to $V_{\vec r_1-\vec r_2,\vec r_3,\vec r_4}\sim\delta^2(\vec r_1-\vec r_2)f_{\vec r_3,\vec r_4}$ for some arbitrary function of $\vec r_3, \vec r_4$. We can again expand the effective bare interaction in the basis of generalized pseudopotentials as given by Eq.(\ref{gppg}). The projected interaction range is defined analogously that $\bm r_c$ is equal to the largest $m$ when $\lambda_{m,n,\vec k_1,\vec k_2}$ is non-zero so that $\lambda_{m>\bm r_c,n,\vec k_1,\vec k_2}=0$. The topological phases realized in these Chern bands are \emph{identical} to the ones realized in the LLs with the same $\bm r_c$, with the only subtlety that the many-body wavefunctions can be characterized by a different guiding center metric $g$ and single particle normalizations; the latter is not important for the physics of interaction\cite{footnote}.

{\it Conformal Hilbert spaces of ideal Chern bands--}While $\bm r_c$ is always finite for a linear combination of a finite number of LLs, in Chern bands this often is no longer the case, as we will discuss in more details later. For any Chern bands that give a finite $\bm r_c$, we term it as the \emph{ideal Chern band}. All generalized LLs\cite{Wang2021, liu2024theorygeneralizedlandaulevels} and vortexable Chern bands\cite{Ledwith2023} described in the literature are ideal Chern bands with finite $\bm r_c$, and it is not clear if they exhaust all possible ideal Chern bands from the perspective of the interaction proposed in this work. Let us first focus on the ideal Chern bands and understand their dynamical properties. The projected interaction range $\bm r_c$ for different Chern bands dictates that different null spaces can be realized, depending on its integer values. These null spaces are spanned by the zero energy states (i.e. the ground states and the quasihole states of the Lauglin phases), termed as the conformal Hilbert spaces due to the emergent conformal symmetry within these null spaces \cite{yang2021gaffnian,Yuzhu2023}. All universal properties of the topological phase are determined by the algebraic properties of such CHS. For example in a Chern band where $\bm r_c=2N$, the highest density CHS that can be exactly realized (bosonic or fermionic) is identical to that of the $N^{\text{th}}$ LL: it is spanned by the Laughlin $\nu=1/\left(2\bm r_c+2\right)$ ground states and quasiholes, with shortest range real space interaction $V(r)=\delta^2(r)$ (and for fermions it would be $\nabla^2\delta^2(r)$). Lower density Laughlin states (both bosonic or fermionic) can be exactly realized with more long range interactions involving $V(r)\sim \nabla^{2n}\delta^{2}(r)$\cite{footnote}. Importantly, Laughlin states with $\nu>1/\left(2\bm r_c+2\right)$ \emph{cannot be exactly} realized in such Chern bands. 

A familiar example is the ideal flat bands of the chiral twisted bilayer graphene (cTBG), the two-body interaction Hamiltonian in the form of Eq.(\ref{cb}) have the following expression
\begin{eqnarray}\label{ctbg}
\tilde V_{\vec q\vec k_1\vec k_2}=&&\tilde V_q\left(1+\sum_{\vec b}w_{\vec b}\left(\mathcal P_{\vec k_1}^{\vec b}e^{-\frac{1}{2}\bm b^*\bm q}+\mathcal P_{\vec k_2-\delta\vec b/2}^{\vec b}e^{\frac{1}{2}\bm b^*\bm q}\right)\right)\nonumber\\
+&&\tilde V_q\left(\sum_{\vec b_1\vec b_2}w_{\vec b_1}w_{\vec b_2}\mathcal Q_{\vec k_1\vec k_2\delta\vec b}^{\vec b_1\vec b_2}e^{\frac{1}{2}\left(\bm b_2^*-\bm b_1^*\right)\bm q}\right)
\end{eqnarray}
where the single particle normalization is ignored. The detailed expressions of $\mathcal P_{\vec k}^{\vec b}$ and $\mathcal Q_{\vec k_1\vec k_2\delta\vec b}^{\vec b_1\vec b_2}$ are shown in the literature\cite{Wang2021,wang2025dynamicslifetimegeometricexcitations}, with $w_{\vec b}$ the parameters of the ideal flat band with vectors $\vec b$ in the momentum space and $w_{\vec b}=w_{-\vec b}$ to ensure hermiticity. Importantly apart from $\tilde V_q$, the effective interaction is holomorphic and only depends on $\bm q$, reflecting the chirality of the band. Just like in the LLs, if $\tilde V_q$ is constant, the pseudopotential expansion only has $\lambda_{0,n,\vec k_1,\vec k_2}$ (or $\lambda_{1,n,\vec k_1,\vec k_2}$) nonzero with $n>0$. It thus has $\bm r_c=0$ for any values of $w_{\vec b}$, not just those values extracted from the cTBG.

We have thus established that for ideal Chern bands with the same $\bm r_c$, the exact CHS of Laughlin phases at $\nu\le 1/\left(2\bm r_c+2\right)$ can be realized with short range TK interaction $\nabla^{2n}\delta^2(r)$. For such Chern bands with short range interactions, not only the topological phases, but also the many-body wavefunctions of the ground states and the quasihole states are exactly identical to those from the (linear combination of) LLs, up to an unimportant single particle normalization factors or a uniformly deformed guiding center metric (also see\cite{footnote}). It is particularly interesting that even with the same physical interaction from Eq.(\ref{hbare}), the projected interaction can be very different from each other with explicit breaking of rotational symmetry, as clearly illustrated by Eq.(\ref{ctbg}). The CHS or the collection of the zero-energy states, however, is completely insensitive to these microscopic details, with full rotational invariance just like those from the LLs with uniform QGT: the fluctuation of the Berry curvature or the QGT plays no role in the physics of the ground states or the gapless edge and bulk quasihole excitations. In particular, topological properties such as the (guiding center) Hall viscosity for the fractional topological phases in these Chern bands are identical to those in the LLs.


The gapped excitations of the topological phases, however, will be strongly affected by the band geometry. These include both the quasielectron and neutral excitations, thus unlike the gapless states of the topological phases, their physical properties can differ significantly from one Chern band to another. In particular unlike the LLs, angular momentum or guiding center spin are no longer good quantum numbers for the gapped excitations. For example neutral geometric excitations within the Chern bands are spin-n excitations constructed from the rank-n tensor describing the area preserving deformation (e.g. the spin-2 gravitons)\cite{Haldane2011,yang2024quantumgeometricfluctuationsfractional,wang2025dynamicslifetimegeometricexcitations}. In generic Chern bands there will be strong scattering between different spins, as strongly evidenced from recent works\cite{wang2025dynamicslifetimegeometricexcitations,paul2025shininglightcollectivemodes}. The dynamics of quasielectrons (which are gapped) will also be more complicated from those of the quasiholes (gapless and identical to those in the LLs apart from the single particle normalization) due to the emergence of rotational symmetry of the latter.

{\it Beyond ideal Chern bands and two-body interactions--}We end this work with some discussions and outlooks. Chern bands in realistic quantum materials are not ideal, though they can be tuned to be close to being ideal\cite{Carr2019,ledwith2020fractional,Dong2023CF}. In these Chern bands, $\bm r_c$ is no longer finite based on the definition in this work, physically implying that the CHS, or model topological phases, cannot be \emph{exactly realized }for any real space electron-electron interaction given in Eq.(\ref{hbare}), in contrast to the LLs. The deviation from the ideal Chern band condition can be treated as a perturbation to the effective bare interaction, analogous to the perturbation to the model pseudopotential interaction in the LLs by, for example, the Coulomb interaction. For incompressible topological phases, if we treat TK interaction in ideal Chern bands as model Hamiltonians, then moving away from TK interaction (e.g. with more realistic electron-electron interaction) or moving away from the ideal Chern band conditions have the similar effect: both can be treated as adiabatic perturbation to the interaction, with the eigenstates of the model Hamiltonians as references, before the incompressibility gap closes. The perturbation thus dresses the CHS including the ground state and the quasihole states but the physics will barely change, as the perturbation is suppressed at least algebraically by the gap. Thus even in realistic systems, topological properties such as the Hall viscosity will not be affected. 

The gapped excitations will however be much more sensitive to the details of the Chern bands. When the effective perturbation is strong, phase transitions will be induced with gap closing. Near the transition the physics of the Chern bands can be very different from those in the LLs, and this is also true after the phase transition, where various symmetry-breaking phases may be realized only in general Chern bands. It will also be interesting to see if there are gapped topological phases that can only be realized in Chern bands other than LLs. Our work shows this implies new topological phases that can only be realized in LLs with \emph{momentum dependent interactions} instead of the density-density interactions, and the analytic tools developed in this work can be useful for such investigations in the future.

The effective interaction within the Chern band from the physical real space interaction given by Eq.(\ref{hbare}) also provides a natural embedding for the lattice systems. It has been argued that the property of the QGT is not fundamentally important for the interacting physics\cite{PhysRevB.102.165148,private}, since different imbedding of the lattices in real space changes the QGT (but not the Chern number). In contrast site based interactions (e.g. on-site or nearest neighbour interaction) and thus the interaction physics are invariant with different embedding. However, with Eq.(\ref{hbare}) the on-site or nearest neighbour interaction is only realized with a particular lattice embedding. For example a nearest neighbour interaction in the honeycomb lattice is realized with a finite range step-function interaction $\tilde V(\vec r)=1$ only for $|\vec r|\le a$, where $a$ is the lattice constant. A different lattice embedding alters both the QGT and the effective interaction of Eq.(\ref{hbare}) projected into the Chern bands. It is thus the special properties of the projected interaction (e.g. with finite $\bm r_c$) that fundamentally dictates the physics, not the QGT or the embedding of the lattice sites in the real space.

While this work only focuses on the two-body interactions, the methodology can also be readily generalized to real space one-body potential or electron-electron interaction involving three or more bodies. The key idea is while in the full Hilbert space it always involves electron density in real space, this is no longer the case in a truncated Hilbert space such as a single (or a group of) band(s). In particular, TK-like few-body interaction in real space also lead to model pseudopotential interactions of a finite range in ideal Chern bands beyond Landau levels, where non-abelian topological phases are realized\cite{PhysRevB.75.075318}. One-body potential projected into a single band leads to effective long-range correlated hopping that is useful in understanding the localization properties of disorders especially in Chern bands. For Chern bands with Chern number $C>1$, we can also understand them as effective momentum dependent interactions within multiple LLs (e.g the bilayer quantum Hall systems). These topics can be studied in more details in future works.

\begin{acknowledgments}
{\sl Acknowledgements.}This work is supported by the National Research Foundation, Singapore under the NRF Fellowship Award (NRF-NRFF12-2020-005), Singapore Ministry of Education (MOE) Academic Research Fund Tier 3 Grant (No. MOE-MOET32023-0003) “Quantum Geometric Advantage”, and Singapore Ministry of Education (MOE) Academic Research Fund Tier 2 Grant (No. MOE-T2EP50124-0017).
\end{acknowledgments}

\bibliography{paper_references}{}

\clearpage

\renewcommand{\thefigure}{S\arabic{figure}}
\renewcommand{\theequation}{S\arabic{equation}}
\renewcommand{\thepage}{S\arabic{page}}
\setcounter{figure}{0}
\setcounter{page}{1}

\onecolumngrid
\begin{center}
\textbf{\large Supplementary Online Materials for ``Equivalence of Chern bands and Landau levels from projected Interactions"}
\end{center}
\setcounter{equation}{0}
\setcounter{figure}{0}
\setcounter{table}{0}
\setcounter{page}{1}
\makeatletter
\renewcommand{\theequation}{S\arabic{equation}}
\renewcommand{\thefigure}{S\arabic{figure}}
\renewcommand{\bibnumfmt}[1]{[S#1]}
\renewcommand{\citenumfont}[1]{S#1}
In this supplementary material, we give more technical details on the analysis of interactions in Chern bands, and how such interactions can be mapped to effective interactions in the Landau levels: flat Chern bands with uniform quantum geometric tensor.

\section{S1. Beyond density-density interaction}\label{s1}

Bare interactions between particles in full Hilbert space are always density-density interaction given by $\hat H_{\text{int}}$ in the form of Eq.(1) in the main text. Depending on the physical context, often we are analysing the dynamics of the interaction with a particular subspace, for example determined by the one-body Hamiltonians (e.g. the kinetic energy, or the one-body potentials) of the system. Here the assumption is that this sub-Hilbert space is gapped from the rest of the Hilbert space, and this gap is the dominant energy scale as compared to the interaction or the temperature.

A simple example is the familiar tight-binding models with the following single particle Hamiltonian
\begin{eqnarray}
\hat H_{\text{TB}}=-\sum_{ij\alpha\beta}t_{i\alpha,j\beta}\hat c^\dagger_{\vec r_{i\alpha}}\hat c_{\vec r_{j\beta}}
\end{eqnarray} 
where $\hat c_{\vec r_{i\alpha}}$ creates a single particle (Wannier) orbital localized at a lattice site with location $\vec r_{i\alpha}$ in real space, with $i$ indexing the unit cell and $\alpha$ indexing the sublattice within the unit cell. It is important to note that physically the associated quantum state created by $\hat c^\dagger_{\vec r_{i\alpha}}$ is not a delta function on $\vec r_{i\alpha}$, but rather a wavefunction $\phi_{i\alpha}\left(\vec r\right)$ localized at $\vec r_{i\alpha}$ with localization length $\ell_a$. The tight-binding model effectively projects into a sub-Hilbert space where every unit cell contributes a fixed number of orthonormal localized orbitals $\phi_{i\alpha}(\vec r)$. One can thus project $\hat H_{\text{int}}$ or Eq.(1) in the main text into this sub-Hilbert space, and the proper microscopic interaction Hamiltonian given by a discrete version of Eq.(2) in the main text with
\begin{eqnarray}\label{generalrd}
&&\hat H=\sum_{ijkl\alpha\beta\gamma\kappa}V_{\vec r_{i\alpha}-\vec r_{j\beta},\vec r_{k\gamma},\vec r_{l\kappa}}\hat c^\dagger_{\vec r_{i\alpha}+\vec r_{k\gamma}}\hat c^\dagger_{\vec r_{j\beta}+\vec r_{l\kappa}}\hat c_{\vec r_{i\alpha}}\hat c_{\vec r_{j\beta}}\quad\\
&&V_{\vec r_1-\vec r_2,\vec r_3,\vec r_4}=\int d^2rd^2r'V\left(\vec r-\vec r'\right)\mathcal F_{\vec r_1\vec r_2\vec r_3\vec r_4}\left(\vec r,\vec r'\right)
\end{eqnarray}
and $\mathcal F_{\vec r_1\vec r_2\vec r_3\vec r_4}\left(\vec r,\vec r'\right)=\phi^*_{\vec r_1}(\vec r')\phi^*_{\vec r_2}(\vec r)\phi_{\vec r_1+\vec r_3}(\vec r')\phi_{\vec r_2+\vec r_4}(\vec r)$. When the density-density interaction in the form of $V_{ij}^{\alpha\beta}\hat n_{i\alpha}\hat n_{j\beta}$ are used (e.g. the nearest neighbour interaction or the Hubbard model, with $\hat n_{i\alpha}=\hat c^\dagger_{i\alpha}\hat c_{i\alpha}$), we are implicitly taking the limit of $\ell_a\to 0$, which is a crude approximation given that $\ell_a$ is generally on the same order as the lattice constant. The effective interaction of Eq.(2) in the main text, or Eq.(\ref{generalrd}) here, can be understood as Eq.(1) in the main text projected into the subspace of an array of selected localized orbitals. Generically speaking, the latter can be thought of as Wannier orbitals from the Bloch states from the periodic one-body potentials in real space.

\section{S2. Short-range real space interactions}

The real-space interaction of Eq.(1) in the main text can also be written as a linear combination of an orthonormal basis of the Trugmann-Kivelson type interaction
 \begin{eqnarray}\label{realpp}
 V_n\left(r\right)\sim\lim_{\ell\to 0}\mathcal L_n\left(\left(r/\ell\right)^{2}\right)e^{-\frac{1}{2}\left(r/\ell\right)^2}
 \end{eqnarray}
 where $\ell$ is the length scale of the interaction range, and $\lim_{\ell\to 0}\mathcal L_n\left(\left(r/\ell\right)^{2}\right)e^{-\frac{1}{2}\left(r/\ell\right)^2}$ is a linear combination of $\nabla^m\delta^{\left(2\right)}\left(r\right)$ with $m\le n$. Without loss of generality we use the shortest range interaction $V_0(r)=\delta^{\left(2\right)}(r)$, one can show that with the corresponding $\tilde V_q=1$ in Eq.(6) in the main text, we have the following pseudopotential expansion with \emph{finite} number of terms:
 \begin{eqnarray}
 \tilde V_q\left(\mathcal L_N\left(q^2/2\right)\right)^2e^{-\frac{1}{2}q^2}=\sum_{m=0}^{N} \lambda_m\tilde V_{2m}(q^2)
 \end{eqnarray}
with $\lambda_m>0$ being readily computable, and crucially $\lambda_m=0$ for $m>N$, which can be proven by simply looking at the power of $q$. Thus with the shortest range interaction in real space, different LLs can be characterized by \emph{different number of pseudopotentials} present in the effective interaction. The higher the LLs, more pseudopotentials will be present and the projected interaction is effectively more ``long ranged". In particular, this implies the exact bosonic Laughlin state at $\nu=1/2$ can only be realised in the LLL, the exact bosonic Laughlin state at $\nu=1/4$ can only be realised in the LLL and the first LL, etc. 
 
This qualitative difference of the behaviour of the real space interaction within different LLs can be generalized to $V_n(r)$ of any non-negative integer $n$. It is an explicit illustration that different LLs cannot be connected by unitary transformations. For fermions the shortest possible physical real space interaction is $V_2(r)$, which is the original Trugman-Kivelson interaction\cite{kivelson1985exact}. Following the arguments above, the null space of the pseudopotential $\tilde V_1(q^2)$, which is spanned by the model ground states and quasihole states of Laughlin $\nu=1/3$ phase, can only be realized in the LLL. Higher LLs can realize the topological phase in the same universality class, but generally a less robust one with a smaller incompressibility gap.

\section{S3. Projection of general interaction Hamiltonians into the Landau levels}

We first briefly show that interaction within the lattice Chern bands can also be completely understood as a generalized real space interaction projected into a Landau level. Starting with a generic lattice tight-binding model with periodic boundary condition:
\begin{eqnarray}
\hat H=-\sum_{i\alpha,j\beta}t_{ij}^{\alpha\beta}\hat c_{\vec r_{i\alpha}}^\dagger c_{\vec r_{j\beta}}+V_{i\alpha,j\beta}\hat n_{\vec r_{i\alpha}}\hat n_{\vec r_{j\beta}}+c.c
\end{eqnarray}
where $\hat c^\dagger_{\vec r_{i\alpha}}$ creates an electron at the $i^{\text{th}}$ unit cell and the $\alpha$ is the sublattice index; $\hat n_{\vec r_{i\alpha}}=\hat c^\dagger_{\vec r_{i\alpha}}\hat c_{\vec r_{i\alpha}}$ is the density operator. This already assumes, as explained in Sec.~\textbf{S1}, that electrons created on the lattice sites are delta functions with no spatial extension. Instead of treating $V_{i\alpha,j\beta}$ as the interaction matrix between different lattice sites with no specific embedding in the real space chosen, we do choose a particular embedding that is commonly assumed, matching the experimental materials the model is intended to describe. In this way, $V_{i\alpha,i\beta}=V_{\vec r_{i\alpha}-\vec r_{j\beta}}$ where $\vec r_{i\alpha}$ is the real space coordinate of the lattice site. For example, the nearest neighbour interaction is given by $V_{\vec r}=\sum_i\delta^{\left(2\right)}\left(\vec r-\vec r_i\right)$, where $\vec r_i$ is a collection of nearest neighbour vectors. On the honeycomb lattice there are six such vectors $\vec r_i\sim\left(0,1\right), \left(\sqrt 3/2,1/2\right), \left(\sqrt 3/2,-1/2\right), \left(0,-1\right), \left(-\sqrt 3/2,-1/2\right), \left(-\sqrt 3/2,1/2\right)$, where we set the lattice constant to be unity. Or one can just take $V_{\vec r}$ to be a rotationally invariant step function with $V_{\vec r}=1$ for $|\vec r|\le 1$ and zero otherwise. The two expressions are completely equivalent for electrons only localized on the lattice sites. 

As usual the kinetic energy part defines the band, giving creation operators of electrons within the $n^{\text{th}}$ band $\psi^\dagger_{\vec k,n}=\sum_{\alpha}u_{\alpha,\vec k,n}\hat d^\dagger_{\vec k,\alpha}$, with $\hat d^\dagger_{\vec k,\alpha}=\sum_ie^{i\vec k\cdot\vec R_{i\alpha}}\hat c^\dagger_{i\alpha}$, where $u_{\alpha,\vec k,n}$ is computed from diagonalizing the fourier component of $t^{\alpha\beta}_{ij}$ in the tight-binding model. The interaction projected into the $n^{\text{th}}$ band is thus given by
\begin{eqnarray}
&&\bar H_{\text{int}}=\sum_{\vec k_1\vec k_2\vec k_3\vec k_4}V_{\vec k_1\vec k_2\vec k_3\vec k_4}\hat d^\dagger_{\vec k_3}\hat d^\dagger_{\vec k_4}\hat d_{\vec k_1}\hat d_{\vec k_2}\label{ccb}\\
&&V_{\vec k_1\vec k_2\vec k_3\vec k_4}=\sum_{\vec b}\tilde V_{\vec q}\sum_{\alpha,\beta}u^*_{\alpha,\vec k_3,n}u^*_{\alpha,\vec k_4,n}u_{\alpha,\vec k_1,n}u_{\alpha,\vec k_2,n}e^{i\vec\delta_{\alpha}\cdot\vec b}e^{-i\vec\delta_{\beta}\cdot\left(\vec b-\delta\vec b\right)}
\end{eqnarray}
with the usual condition that $\vec k_i$ are the crystal momentum within the first Brillouin zone with $\vec k_1+\vec k_2=\vec k_3+\vec k_4+\delta\vec b$ from discrete translational symmetry, and $\vec q=\vec k_1-\vec k_4-\vec b$ the momentum transfer. Both $\vec b$ and $\delta\vec b$ are the reciprocal vectors. Note here $\vec \delta_{\alpha}$ is the relative position of the $\alpha^{\text{th}}$ sublattice within the unit cell. 

From this formulation, the interaction Hamiltonian for the Chern bands in the lattice system has the same form as those of the Landau levels or the Moire systems, with $\tilde V_{\vec q}$ the fourier component of $V_{\vec r}$. Both from a physical and especially from a numerical point of view, different Chern bands have exactly the same Hilbert space, and are only distinguished by different single particle form factors, and the reciprocal vectors containing the information about the lattice type. The physical external interaction is always given by $\tilde V_{\vec q}$, and in Landau levels, the lattice type can be arbitrarily chosen as long as the number of ``lattice sites" is equal to the number of orbitals in a single Landau level. 

Thus Eq.(\ref{ccb}) can be understood as the equivalent to the following general interaction Hamiltonian
\begin{eqnarray}\label{unprojected}
\hat H_{\text{eff}}=\sum_{\vec k_1\vec k_2\vec k_3\vec k_4}\bar V_{\vec k_1\vec k_2\vec k_3\vec k_4}\hat d^\dagger_{\vec k_3}\hat d^\dagger_{\vec k_4}\hat d_{\vec k_1}\hat d_{\vec k_2}
\end{eqnarray}
projected into the LLL, where $\bar V_{\vec k_1\vec k_2\vec k_3\vec k_4}=V_{\vec k_1\vec k_2\vec k_3\vec k_4}\left(\mathcal F^{\text{LLL}}_{\vec k_1\vec k_2\vec k_3\vec k_4}\right)^{-1}$ as detailed in the main text, where $d^\dagger_{\vec k}$ creates a plane wave $\sim e^{i\vec k\cdot\vec r}$ in the full Hilbert space. After projecting into the LLL we have the following:
\begin{eqnarray}\label{projected}
\bm P_{\text{LLL}}\hat H_{\text{eff}}\bm P_{\text{LLL}}=\sum_{\vec q_1\vec q_2\vec q_3\vec q_4}\sum_{\vec b_1\vec b_2\vec b_3\vec b_4}\bar V_{\vec q_1+\vec b_1,\vec q_2+\vec b_2, \vec q_3+\vec b_3, \vec q_4+\vec b_4}\left(\mathcal F^{\text{LLL}}_{\vec q_1+\vec b_1,\vec q_2+\vec b_2, \vec q_3+\vec b_3, \vec q_4+\vec b_4}\right)^{-1}\hat \psi^\dagger_{\vec q_3}\hat \psi^\dagger_{\vec q_4}\hat \psi_{\vec q_1}\hat \psi_{\vec q_2}
\end{eqnarray}
where $\vec b_i$ are the reciprocal vectors.

Note that in the full Hilbert space $\vec k$ can be any wave vectors, while after the projection into the sub-Hilbert space, $\vec q$ is restricted to the first Brillouin zone with discrete translational invariance and thus $\vec q_1+\vec q_2-\vec q_3-\vec q_4=\delta \vec b$ with some reciprocal vector $\delta \vec b$. The summation over $\vec b_i$ in Eq.(\ref{projected}) comes from the fact that for a plane wave state $|\vec k\rangle$ and a Bloch state $|\vec q\rangle$, we have $\langle \vec k|\vec q\rangle=\sum_{\vec b}\delta^2\left(\vec k-\vec q+\vec b\right)$. Naturally for any periodic systems, we would require $\bar V_{\vec q_1+\vec b_1,\vec q_2+\vec b_2, \vec q_3+\vec b_3, \vec q_4+\vec b_4}=\bar V_{\vec q_1\vec q_2\vec q_3\vec q_4}$ for any physical interaction. Here the LLL is spanned by the Bloch states, so it is the ideal flat band in a periodic system with no magnetic field, e.g. the $w_{\vec b_1}=0$ flat band given in Ref.\cite{Wang2021}. 

Since Eq.(\ref{unprojected}) is the interaction Hamiltonian in full Hilbert space, we can also transform it into the real space basis. Translational invariance requires $\bar V_{\vec k_1\vec k_2\vec k_3\vec k_4}$ to only depend on $\vec k_3,\vec k_4$, $\vec k_1-\vec k_3$ which is equal to $\vec k_4-\vec k_2$. In real space the equivalent interaction is thus given by:
\begin{eqnarray}
\hat H^{(r)}_{\text{eff}}=\int d^2\vec r_1d^2\vec r_2d^2\vec r_3d^2\vec r_4\tilde V_{\vec r_1-\vec r_2,\vec r_3,\vec r_4}\hat c^\dagger_{\vec r_1+\vec r_3}\hat c^\dagger_{\vec r_2+\vec r_4}\hat c_{\vec r_1}\hat c_{\vec r_2}
\end{eqnarray}
with $\tilde V_{\vec r_1\vec r_2\vec r_3}=\int d^2\vec k_1d^2\vec k_2d^2\vec k_3e^{i\vec k_1\cdot\vec r_1}e^{i\vec k_2\cdot\vec r_2}e^{i\vec k_3\cdot\vec r_3}\bar V_{\vec k_2+\vec k_1,\vec k_3-\vec k_1,\vec k_2,\vec k_3}$. It is now more convenient to see the difference between the projection of full Hilbert space Hamiltonians into the LLL with the Bloch states $|\vec q\rangle$ (without the magnetic field) or the magnetic Bloch states $|\tilde q\rangle$ (with the magnetic field). In both cases $\vec q,\tilde q$ are wave vectors within the first Brillouin zone. In real space, we can write the magnetic Bloch states in the following way:
\begin{eqnarray}\label{realspace}
\tilde\psi_{\tilde q}\left(\vec r\right)=e^{i\int_0^{\vec r}d\vec l\cdot\vec A}\psi_{\tilde q}(\vec r)
\end{eqnarray}
where $\vec A$ is the vector potential that gives a uniform magnetic field, and the line integral can be taken as the straight line from the origin to $\vec r$. While $\tilde\psi_{\tilde q}(\vec r)$ is quasiperiodic due to the vector potential, $\psi_{\tilde q}(\vec r)$ is the periodic Bloch state. Thus projection of Eq.(\ref{realspace}) into the LLL with Bloch states (no magnetic field) will give Eq.(\ref{projected}); but when projecting into the LLL with the magnetic field, we need to replace $\tilde V_{\vec r_1-\vec r_2,\vec r_3,\vec r_4}$ with $\tilde V_{\vec r_1-\vec r_2,\vec r_3,\vec r_4}e^{i\int_0^{\vec r_1}d\vec l\cdot\vec A}e^{i\int_0^{\vec r_2}d\vec l\cdot\vec A}e^{-i\int_0^{\vec r_2+\vec r_4}d\vec l\cdot\vec A}e^{-i\int_0^{\vec r_1+\vec r_3}d\vec l\cdot\vec A}$. Though the additional phase is gauge dependent, the energy spectrum of the interaction will not depend on the gauge choice.

\section{S4. The formalism of momentum dependent generalized pseudopotentials}

For any two-body interaction in real space, its fourier component in momentum space can be expanded in a complete orthonormal basis as follows\cite{yang2017generalized}:
\begin{eqnarray}
\tilde V_{\vec q}=\sum_{m,n}\lambda^{\left(1\right)}_{m,n}\tilde V_{m,n}\left(\bm q, \bm q^*\right)+\lambda^{\left(2\right)}_{m,n}\tilde V_{m,n}\left(\bm q^*, \bm q\right)
\end{eqnarray}
where $\tilde V_{m,n}\left(\bm q, \bm q^*\right)=\sqrt{\frac{m!}{\pi\left(m+n\right)!}}\mathcal L_{m}^n\left(q^2\right)\bm q^n$ with $\mathcal L(x)$ being the generalized Laguerre polynomials, and constraint that $m,n\ge 0$. The two-body operator
\begin{eqnarray}
\hat O=\int d^2q \tilde V_{m,n}\left(\bm q, \bm q^*\right)\hat c^\dagger_{\vec k_1+\vec q}\hat c^\dagger_{\vec k_2-\vec q}\hat c_{\vec k_1}\hat c_{\vec k_2}
\end{eqnarray}
maps a pair of electrons with relative angular momentum $m$ to the relative angular momentum $m+n$, while keeping the center of mass angular momentum intact. Thus only even values of $n$ are relevant due to particle statistics, and for the Hamiltonian to be Hermitian, we need to have $\lambda^{\left(1\right)}_{m,n}=\left(\lambda^{\left(2\right)}_{m,n}\right)^*$. These coefficients can be computed by
\begin{eqnarray}
&&\lambda^{\left(1\right)}_{m,n}=\int d^2q\tilde V_{\vec q}\tilde V_{m,n}\left(\bm q,\bm q^*\right)e^{-\frac{1}{2}q^2}\\
&&\lambda^{\left(2\right)}_{m,n}=\int d^2q\tilde V_{\vec q}\tilde V_{m,n}\left(\bm q^*,\bm q\right)e^{-\frac{1}{2}q^2}
\end{eqnarray}
for $n>0$. Note that the usual rotationally invariant Haldane pseudopotentials are special cases with $n=0$ thus $\lambda^{\left(1\right)}_{m,0}=\lambda^{\left(2\right)}_{m,0}$, thus an additional factor of $1/2$ is needed. 

For a generic interaction $\tilde V_{\vec q,\vec k_1,\vec k_2}$, we can also define the following:
\begin{eqnarray}
&&\lambda^{\left(1\right)}_{m,n,\vec k_1,\vec k_2}=\int d^2q\tilde V_{\vec q,\vec k_1,\vec k_2}\tilde V_{m,n}\left(\bm q,\bm q^*\right)e^{-\frac{1}{2}q^2}\\
&&\lambda^{\left(2\right)}_{m,n,\vec k_1,\vec k_2}=\int d^2q\tilde V_{\vec q,\vec k_1,\vec k_2}\tilde V_{m,n}\left(\bm q^*,\bm q\right)e^{-\frac{1}{2}q^2}
\end{eqnarray}
We can thus expand the generic interaction with the same basis, but the coefficients now are momentum dependent:
\begin{eqnarray}
\tilde V_{\vec q,\vec k_1,\vec k_2}=\sum_{m,n}\lambda^{\left(1\right)}_{m,n,\vec k_1,\vec k_2}\tilde V_{m,n}\left(\bm q, \bm q^*\right)+\lambda^{\left(2\right)}_{m,n,\vec k_1,\vec k_2}\tilde V_{m,n}\left(\bm q^*, \bm q\right)
\end{eqnarray}
It is actually more natural to use the relative and center of mass linear momentum with $\vec k = \vec k_1-\vec k_2,\vec K=\vec k_1+\vec k_2$, so the condition of Hermiticity requires
\begin{eqnarray}
\tilde V_{\vec q,\vec k,\vec K}=\tilde V_{-\vec q,\vec k+2\vec q,\vec K}^*
\end{eqnarray}
Interestingly with the generic interaction which is physically well motivated Sec.~\textbf{S1} as well as in the main text, $\tilde V_{\vec q,\vec k,\vec K}$ can be completely holomorphic or anti-holomorphic in $\vec q$. This is the main reason ideal Chern bands can be realized beyond the Landau levels, as we detailed in the main text. A general form of $\tilde V_{\vec q, \vec k,\vec K}$ that is holomorphic in $\vec q$ is
\begin{eqnarray}
&&\tilde V_{\vec q,\vec k,\vec K}=\tilde V_{\vec q}\left(f_{\vec K}g_{\vec k_{\bm q}}-f^*_{\vec K}g_{-\vec k_{\bm q}}\right)\\
&&\vec k_{\bm q}=\bm \alpha^*\bm q+i\vec k\times \vec\alpha
\end{eqnarray}
where $\tilde V_{\vec q}$ is from the bare real space interaction, $f,g$ are arbitrary functions but $g$ is a real function, $\vec\alpha =\alpha_x+i\alpha_y$ is an arbitrary vector with $\bm\alpha=\alpha_x+i\alpha_y$. In particular, any Chern band equivalent to such an effective interaction is an ideal Chern band with $\bm r_c=0$. 

Just like the conventional pseudopotentials where there is a metric degree of freedom\cite{yang2012band,yang2017generalized}, the momentum dependent generalized pseudopotentials can also be parameterized by a unimodular metric $g^{ab}$ which defines the inner product $q_g^2=\bm q_g\bm q_g^*=g^{ab}q_aq_b$. This metric is called the ``guiding center metric". With this geometric degree of freedom, the same interaction $\tilde V_{\vec q,\vec k_1,\vec k_2}$ can have coefficients of expansion dependent on the metric as below:
\begin{eqnarray}
&&\lambda^{\left(1\right),g}_{m,n,\vec k_1,\vec k_2}=\int d^2q\tilde V_{\vec q,\vec k_1,\vec k_2}\tilde V_{m,n}\left(\bm q_g,\bm q_g^*\right)e^{-\frac{1}{2}q_g^2}\\
&&\lambda^{\left(2\right),g}_{m,n,\vec k_1,\vec k_2}=\int d^2q\tilde V_{\vec q,\vec k_1,\vec k_2}\tilde V_{m,n}\left(\bm q_g^*,\bm q_g\right)e^{-\frac{1}{2}q_g^2}
\end{eqnarray}
For some systems where the intrinsic metric is non-trivial, one needs to tune the guiding center metric to find the ``optimal" metric such that $\bm r_c$ is the smallest. This is because even if $\bm r_c=0$ for a specific choice of $g$, all $\lambda_{m,n,\vec k_1,\vec k_2}^{\left(i\right), g'}$ could be nonzero for $g'\ne g$. Determining the optimal $g$ is not easy, and an alternative, more practical method is to look at the bosonic ground state energy at filling factor $\nu=1/(2p+2)$ with $\tilde V_q=1$. Now $\bm r_c$ can be determined by equaling it to the smallest $p$ that gives exact zero ground state energy, corresponding to the bosonic Laughlin state.

\bibliography{paper_references}{}

\end{document}